\documentclass{new_tlp}
\usepackage{mathptmx}
\usepackage{xcolor}
\usepackage{url}
\usepackage{hyperref}
\usepackage{amsmath}
\usepackage{amsfonts}

\usepackage{tikz}
\usepackage{tkz-graph}
\usetikzlibrary{snakes,shapes}
\usetikzlibrary{calc,shapes.multipart,chains}
\usetikzlibrary{arrows,automata}
\usetikzlibrary{positioning}

\usepackage{algorithm} 		
\usepackage{algpseudocode} 	

\newcommand{\TablePaths}{\mathit{TablePaths}}
\newcommand{\TableProb}{\mathit{TableProb}}

\newcommand{\comp}{\mathit{comp}}
\newcommand{\Res}{\mathit{Res}}

\newtheorem{example}{Example} 

\title[Optimizing Probabilities in Probabilistic Logic Programs] 
{Optimizing Probabilities in Probabilistic Logic Programs}

\author[Damiano Azzolini and Fabrizio Riguzzi] {
  DAMIANO AZZOLINI \\
  Dipartimento di Ingegneria - University of Ferrara, Via Saragat 1, I-44122, Ferrara, Italy \\
  \email{damiano.azzolini@unife.it}
  \and
  FABRIZIO RIGUZZI \\
  Dipartimento di Matematica e Informatica - University of Ferrara, Via Saragat 1, I-44122, Ferrara, Italy \\
  \email{fabrizio.riguzzi@unife.it}
}

\jdate{March 2003}
\pubyear{2003}

\newtheorem{definition}{Definition} 

\DeclareMathOperator*{\argmin}{arg\,min}

\begin{document}


\maketitle

\begin{abstract}
Probabilistic Logic Programming is an effective formalism for encoding problems characterized by uncertainty.
Some of these problems may require the optimization of probability values subject to constraints among probability distributions of random variables.
Here, we introduce a new class of probabilistic logic programs, namely Probabilistic \textit{Optimizable} Logic Programs, and we provide an effective algorithm to find the best assignment to probabilities of random variables, such that a set of constraints is satisfied and an objective function is optimized. This paper is under consideration for acceptance in Theory and Practice of Logic Programming.

\end{abstract}

    
\section{Introduction}
\label{sec:intro}
Uncertainty on the data is ubiquitous and pervades almost all domains, such as social networks (uncertainty on the relations), power networks (uncertainty on the components), and road networks (uncertain about the traffic). 
Many real-world applications that can be modelled with random variables require the solution of optimization problems.
In particular, considering the domains listed above, in a social network or in a collaboration network we may want to optimize the probability of targeting some people (or a group of them) during a marketing campaign, but we may be uncertain whether the people (or group of people) know each other, so we need to introduce random variables to model the data. 
In road networks we may want to minimize the probability of encountering a traffic jam, but typically we are unsure about the traffic distribution, so probabilities are needed. 
In power networks we can, for example, optimize the positioning of several electronic components to reduce the probability of a system fault, given that these components have a certain tolerance and a performance degradation after several running hours.
Furthermore, uncertainty is often an intrinsic characteristic of the data itself (collected by faulty sources, incomplete, \dots), so probabilities are necessary.
The same type of task can as well be of interest for other, more theoretical, domains; just to name a few: in Markov network we can set the probability of the possible transitions to optimize the chances of reaching a certain final state, in probabilistic context-free grammars we may be interested in finding the probability of the various components such that the classification of a sentence does not change.

Probabilistic Logic Programming~\cite{DBLP:conf/ilp/2008p,Rig18-BK} is an extension of Logic Programming that allows to encode probabilistic models with a rich language. 
In this paper, we address the problem of optimizing an objective function by tuning the probabilities of probabilistic logic programs subject to constraints on probabilities of facts and queries.

To solve the aforementioned task, we extend the PITA reasoner~\cite{RigSwi10-ICLP10-IC} to allow the definition of \textit{optimizable} facts, i.e., facts whose probabilities can be set in order to optimize (minimize or maximize) an objective function subject to constraints.
Constraints can be on both the query and the optimizable facts.
We introduce the definition of Probabilistic Optimization Logic Program and Probabilistic Optimizable Problem, and provide an algorithm to solve it.
This task can be considered as constrained parameter learning.

By extending the library PITA, we can express models that retain the full LPAD expressive power, while adding new optimizable facts: in this way, we do not need to rewrite a specialized system to solve the problem, we just need to build upon the previous developed work.

The paper is organized as follows: Section~\ref{sec:related} analyzes the related works, Section~\ref{sec:plp} presents the basic definitions regarding Probabilistic Logic Programming, and Section~\ref{sec:polp} introduces the definition of the task we consider, together with an algorithm to solve it.
Section~\ref{sec:exp} presents some empirical results and Section~\ref{sec:concl} concludes the paper. 

\section{Related Works}
\label{sec:related}
A huge part of related works focuses on parameter learning, i.e., learning the parameters of a probabilistic logic program in a supervised manner, starting from a set of examples. 
However, explicit constraints on probabilities of probabilistic facts are usually not considered. 
Here, we learn the parameter (probabilities) of the program's facts starting from a set of constraints, so our algorithm does not need a training set, since it can be classified as an algorithm for solving constrained optimization problems.

One of the first approaches to solve the parameter learning task can be found in PRISM~\cite{DBLP:conf/iclp/Sato95}: the goal is to find the maximum likelihood parameters of special atoms, called $msw$ atoms, that are not present in the dataset.
To learn the parameters, they propose a naive implementation of the EM algorithm, later improved in~\cite{DBLP:journals/jair/SatoK01}.
EM is also used in~\cite{BelRig12-ILP11-IC} to learn the parameters of Logic Programs with Annotated Disjunctions (LPADs)~\cite{VenVer04-ICLP04-IC} represented as Binary Decision Diagrams (BDDs).
Another technique is presented in ProbLog2~\cite{DBLP:journals/corr/abs-1304-6810} that includes the algorithm LFIProbLog~\cite{DBLP:conf/pkdd/GutmannTR11}, where parameters of ProbLog programs are learned from partial interpretations.

Parameter learning using gradient-based methods is applied in~\cite{DBLP:conf/pkdd/GutmannKKR08}, where derivatives are computed directly on the BDD representation of the program.
Here, we use gradient-based methods as well (see Section~\ref{sec:exp}), but we do not compute gradients directly on the BDD. 
Rather, we obtain the equation represented by the BDD and then use this equation (after simplification) in all the calculations.
In this way, we traverse the BDD only once.
A related idea is introduced in~\cite{kimmig2011algebraic} where the authors present aProbLog, an extension of the programming language ProbLog~\cite{raedt07problog}.
It allows to solve different tasks, among them, the sensitivity analysis, i.e., see how a change in the parameters affects the probability of a query.
However, they cite the task in passing, and they do not consider further constraints on the probability of the variables. 
The same problem is discussed in~\cite{orsini2017kproblog}.
Other approaches, such as DTProblog~\cite{DBLP:conf/aaai/BroeckTOR10}, extend probabilistic logic programs with Boolean decision variables. The goal is to find the optimal set of these variables to maximize the overall expected utility. As happens for previously described approaches, no constraints are considered.

Another related solution is presented in~\cite{10.1007/978-3-319-66158-2_32}, where the authors introduce an algorithm to combine Probabilistic (Logic) Programming and Constraint Programming to solve decision-theoretic tasks: as in that paper, we use a compact representation of the probabilistic logic program (they use SDD, we use BDD), and extend an already existing tool (they extend ProbLog, we extend PITA). 
Differently, they restrict the type of constraints involved (linear constraints over sum of Boolean decision variables) and they do not consider the computation of optimal probability values.

A research area that combines probability and constraints is stochastic constraint programming (SCP)~\cite{walsh2002}. 
SCP programs are composed of two types of variables: decision variables that can be set, and stochastic variables that follow a probability distribution.
The goal is to find a subset of decision variables such that constraints are satisfied with a certain probability.
Here, we do not consider probabilistic constraints, we have hard constraints that must be always true, and we search for optimal probability values for optimizable facts that are present in the program (they cannot be removed, as happens for decision variables) but with a tunable probability. Furthermore, these approaches are often tailored to solve specific problems such as games~\cite{antuori2019iee}, scheduling~\cite{lombardi2010aij}, or sequential planning~\cite{babaki2017ijcai}.

\section{Probabilistic Logic Programming}
\label{sec:plp}
Logic Programming offers the possibility to encode problems with a high level of abstraction, allowing programmers to focus on the overall structure, rather than on implementation details.
However, logic programs are not well suited to handle uncertainty, an intrinsic feature of real world scenarios.
Probabilistic Logic Programming (PLP)~\cite{DBLP:conf/ilp/2008p,Rig18-BK} extends Logic Programming by adding probabilistic facts that allow to reason with probabilities.
Following the ProbLog~\cite{raedt07problog} syntax, a probabilistic fact $f_i$ has the form
$$
\Pi_i::f_i
$$
where $f_i$ is an atom and $\Pi_i \in ]0,1]$, with the meaning that each ground instantiation of $f_i$ is true with probability $\Pi_i$ and false with probability $1 - \Pi_i$.
Probabilistic facts are considered independent: this may seem a severe restriction but, in practice, does not limit the expressiveness of the language~\cite{Rig18-BK}.
If we consider the following illustrative example:
\begin{verbatim}
0.5::no_traffic.
0.9::no_accidents.

on_time:- no_traffic, no_accidents.
\end{verbatim}
we can ask the probability that a driver will arrive on time at work (\texttt{on\_time}), which is $0.5\cdot0.9 = 0.45$.
Similarly, An LPAD is composed of a finite set of probabilistic disjunctive clauses of the form
$$h_{1}:\Pi_{1}; \ldots ; h_{n}:\Pi_{n} :- b_{1},\ldots,b_{m}.$$
where the semicolon stands for disjunction, $h_{1},\ldots h_{n}$ are logical atoms and $b_{1},\ldots ,b_{m}$ are logical literals. 
$\Pi_{1},\ldots,\Pi_{n}$ are real numbers in the interval $[0,1]$ that sum to $1$.
If this is not true, the head of the annotated disjunctive clause implicitly contains an extra atom $null$ that does not appear in the body of any clause; its annotation is $1 - \sum_{k=1}^{n} \Pi_{k}$.

One of the most followed semantics, the Distribution Semantics~\cite{DBLP:conf/iclp/Sato95}, is at the foundations of several PLP systems.
Here, we consider the Distribution Semantics for programs with a finite Herbrand base (without function symbols).
An \emph{atomic choice} is a selection or not of a grounding $f\theta$ of a probabilistic fact $\Pi :: f$. 
It is usually indicated with $(f,\theta,s)$, where $s \in \{0,1\}$: if $s = 1$ the fact is selected, if $s = 0$ it is not.
A set of atomic choices is \emph{consistent} if only one of the two alternatives for a probabilistic fact is considered, that is, it does not contain two atomic choices $(f,\theta,s)$ and $(f,\theta,t)$ with $s \neq t$.
A consistent set of atomic choices forms a \emph{composite choice}, and if it contains one atomic choice for every grounding of each probabilistic fact it is called \emph{selection}.
The probability of a composite choice $c$ can be computed as: $$
P(c) = \prod_{(f_i,\theta,1) \in c} \Pi_i \cdot \prod_{(f_i,\theta,0) \in c} (1-\Pi_i) $$
A selection identifies a \emph{world} (a Logic Program) and its probability can be computed as the product of the probabilities of the atomic choices.
The programs do not contain function symbols, thus the set of worlds is finite and the probabilities of all the worlds add up to 1.
Given a ground query $q$, the conditional probability of $q$ given a world $w$, represented with $P(q \mid w)$, is 1 if the query is true in the world, 0 otherwise.
So, the probability of a (ground) query $q$ can be obtained marginalizing the joint probability distribution between the worlds $w$ and the query, $P(q,w)$. 
In formulas:
\begin{equation*}
P(q) = \sum_{w} P(q,w) = \sum_{w} P(q \mid w) \cdot P(w) = \sum_{w \models q} P(w)
\end{equation*}
A composite choice is an \textit{explanation} for a query $q$ (a conjunction of ground atoms) if $q$ is true in every world identified by the composite choice. 
If every world in which $q$ is true belongs to the worlds identified by a set of explanations, the set is named \emph{covering}.
The probability of a query can also be defined as the probability measure of a covering set of explanations.

A covering set of explanations for a query can be represented as a Boolean formula in Disjunctive Normal Form (DNF). 
To perform inference on these formulas, usually probabilistic logic programs are compiled into a target language using knowledge compilation~\cite{DBLP:journals/jair/DarwicheM02}.
Among all the possibilities, Binary Decision Diagrams (BDDs) and Sentential Decision Diagrams (SDDs) are the most used, since they offer a compact representation. Between the two, SDDs are more compact than BDDs. 
However, BDDs are managed by optimized libraries and usually offer comparable performance.
In this paper, we consider BDDs.

A BDD is a direct acyclic graph where each node (that in the PLP setting represents a probabilistic fact) has two edges: the 0 (false) edge and the 1 (true) edge.
There are only two types of terminal nodes: 0 (false) and 1 (true). 
Some packages introduce the possibility to have a third edge, the 0-complemented edge, with the meaning that the function represented by the subtree must be complemented. Here we use this convention.
With this third type, the 0-terminal is not needed.
Usually, the 0-edge is graphically represented with a dashed line, the 1-edge with a solid line and the 0-complemented edge with a dotted line (see Fig.~\ref{fig:bdd_prog} with the running example used in the paper that will be discussed later).

\section{Probabilistic Optimizable Logic Programs}
\label{sec:polp}
We denote \textit{optimizable} facts with the special functor \verb|optimizable|. They have the following syntax: 
$$
\mathrm{optimizable\ } [\Pi_{lb},\Pi_{ub}]::a.
$$
where $a$ is a logical term, and $\Pi_{lb}$ and $\Pi_{ub}$ are the lower and upper probabilities for the term ($\Pi_{lb} < \Pi_{ub}$).
In our implementation, if lower and upper probabilities are not specified, the range is by default $[0.001,0.999]$.
Intuitively, these are the facts whose probabilities can be changed to minimize an objective function subject to constraints. 
Both the objective function (that may not be the same of the query) and the constraints can be linear or non-linear combinations of optimizable facts. 
As happens for probabilistic facts, also optimizable facts are considered independent. 
We now introduce the definition of \textit{Probabilistic Optimizable Logic Program} (POLP) and \textit{Probabilistic Optimizable Problem}.
\begin{definition}[Probabilistic Optimizable Logic Program]
\label{def:polp}
Given an LPAD $\mathcal{L}$, a set of optimizable facts $\mathcal{O}$, an objective function $\mathcal{F}$, and a set of constraints $\mathcal{C}$, the tuple $(\mathcal{L},\mathcal{O},\mathcal{F},\mathcal{C})$ identifies a \textit{Probabilistic Optimizable Logic Program} (POLP).
\end{definition}

\begin{definition}[Probabilistic Optimizable Problem]
\label{def:popl_obj}   
Given a Probabilistic Optimizable Logic Program $(\mathcal{L},\mathcal{O},\mathcal{F},\mathcal{C})$, and a conjunction of ground atoms, the \textit{query} ($q$), the probabilistic optimizable problem consists of two steps:
\begin{itemize}
\item find a probability assignment $\mathcal{A^*}$ to optimizable facts $o_i \in \mathcal{O}$ such that the objective function is minimized (or maximized) and constraints are not violated, i.e.: $$ \mathcal{A}^* = \argmin_{\mathcal{A}} (\mathcal{F} \mid \mathcal{C}, \mathcal{A})$$
\item compute the probability of the query given these assignments $$ P(q \mid \mathcal{A}^*) $$ 
\end{itemize}
This task can also be considered as parameter learning under constraints.
\end{definition}

\begin{figure}[ht]
\begin{minipage}{\linewidth}
\begin{minipage}{0.45\textwidth}
\begin{verbatim}
0.9::edge(a,b).
optimizable [0.3,0.8]::edge(b,c).
optimizable [0.3,0.8]::edge(b,d).
0.3::edge(c,e).
0.8::edge(d,e).

path(X,X).
path(X,Y):- path(X,Z), edge(Z,Y).
\end{verbatim}
\end{minipage}
\hfill
\begin{minipage}{0.45\textwidth}
\centering
\begin{tikzpicture}[->,>=stealth',shorten >=1pt,auto,node distance=1.5cm,semithick,scale = 0.5]
\tikzstyle{every state}=[fill=white,draw=black,thick,text=black,scale=0.8]

\node[state] (A)              {\texttt{a}};
\node[state] (B) [right of=A] {\texttt{b}};
\node[state] (C) [above right of=B] {\texttt{c}};
\node[state] (D) [below right of=B] {\texttt{d}};
\node[state] (E) [below right of=C] {\texttt{e}};

\path (A) edge 				node {}	(B)
      (B) edge 				node {}	(C)
      (B) edge 				node {}	(D)
      (C) edge 				node {}	(E)
      (D) edge 				node {}	(E);
\end{tikzpicture}

\end{minipage}
\end{minipage}
\caption{Program for motivating example (left), together with the represented graph (right).}
\label{fig:program}
\end{figure}
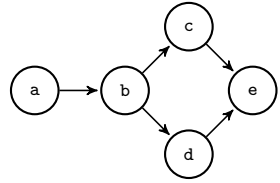

Consider the following motivating example:
\begin{example}
\label{ex:motivating}
Suppose you need to route a signal through a path composed of intermittent edges (connections), from a source to a given destination.
Some of the edges have a fixed probability to be active, while some other edges can be controlled, and their probabilities can be set. 
The constraint is that the probability to reach the destination must be above a certain threshold.
The objective function is that the probability of the edges that can be controlled should be set at the minimal value to reach the target probability, since setting higher probabilities also requires higher manufacturing costs. 
Moreover, the probabilities of the edges should be similar (i.e., as close as possible).
\end{example}
Let us model the previous scenario with a Probabilistic Optimizable Logic Program.
Consider five nodes named \verb|a|, \verb|b|, \verb|c|, \verb|d| and \verb|e|. 
Their connections can be represented with \verb|edge/2| facts.
Suppose that edges between \verb|a| and \verb|b|, \verb|c| and \verb|e|, and \verb|d| and \verb|e| cannot be controlled and have, respectively, probability $0.9$, $0.3$ and $0.8$. 
Edges between \verb|b| and \verb|c| and \verb|b| and \verb|d| can be controlled and must have probability in the range $[0.3,0.8]$.
A path between two nodes is represented by predicate \verb|path/2| (see Fig.~\ref{fig:program}).
The goal is to minimize the sum of the probabilities of optimizable variables (\verb|edge(b,c)| and \verb|edge(b,d)|) given that the probability to reach \verb|e| from \verb|a| (\verb|path(a,e)|) must be above a certain threshold.
Constraints also impose that the difference between the probabilities of the two optimizable facts should be less than, for example, 0.1.
The POLP shown in Fig.~\ref{fig:program} represents the described situation.

To simplify the notation, in the remaining part of the paper we remove the explicit probability signature from facts involved in the optimization task. 
For example, $edge(b,c)$ represents the probability of $edge(b,c)$ ($P(edge(b,c))$).
If the goal is to minimize the sum of the probabilities of \verb|edge(b,c)| and \verb|edge(b,d)|, the lower bound probability for query \verb|path(a,d)| is 0.6, and the difference between the probabilities of the optimizable facts should be less than 0.1, following Def.~\ref{def:polp} we get: 
\begin{itemize}
    \item $\mathcal{O} = \{edge(b,c),edge(b,d)\}$ 
    \item $\mathcal{F} = minimize(edge(b,d) + edge(b,c))$ 
    \item $\mathcal{C} = \{path(a,d) > 0.6, edge(b,d) \in [0.3,0.8], edge(b,c) \in [0.3,0.8], \\ edge(b,c) - edge(b,d) < 0.1, edge(b,d) - edge(b,c) < 0.1\}$
\end{itemize}
Implicitly, $path(a,d) \in [0,1]$.

To solve the probabilistic optimizable problem, we introduce the special predicate 
\begin{verbatim}
prob_optimize/4 
\end{verbatim}
that receives as input the query (in our example \verb|path(a,e)|), the objective function to be minimized (\verb|edge(b,c) + edge(b,d)|), and a list of constraints. 
As output, it returns the probability of the query and the optimal probability assignment to optimizable facts. 
So, for the previous example, the query would be 
\begin{verbatim}
prob_optimize(
    path(a,e),
    [edge(b,c) + edge(b,d)], 
    [path(a,e) > 0.6, edge(b,c) - edge(b,d) < 0.1, 
     edge(b,d) - edge(b,c) < 0.1], 
    Assignments).   
\end{verbatim}

Lower and upper bounds are directly considered by the predicate, without the need to be specified also in the constraints list.

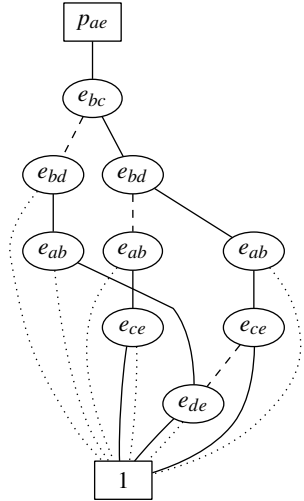
\begin{figure}
\centering
\begin{tikzpicture}[>=latex',line join=bevel,scale=0.4]
    \pgfsetlinewidth{0.5bp}
      \begin{scope}
        \pgfsetstrokecolor{black}
        \definecolor{strokecol}{rgb}{1.0,1.0,1.0};
        \pgfsetstrokecolor{strokecol}
        \definecolor{fillcol}{rgb}{1.0,1.0,1.0};
        \pgfsetfillcolor{fillcol}
        \filldraw (0.0bp,0.0bp) -- (0.0bp,468.0bp) -- (401.38bp,468.0bp) -- (401.38bp,0.0bp) -- cycle;
      \end{scope}
      \begin{scope}
        \pgfsetstrokecolor{black}
        \definecolor{strokecol}{rgb}{1.0,1.0,1.0};
        \pgfsetstrokecolor{strokecol}
        \definecolor{fillcol}{rgb}{1.0,1.0,1.0};
        \pgfsetfillcolor{fillcol}
        \filldraw (0.0bp,0.0bp) -- (0.0bp,468.0bp) -- (401.38bp,468.0bp) -- (401.38bp,0.0bp) -- cycle;
      \end{scope}
      \begin{scope}
        \pgfsetstrokecolor{black}
        \definecolor{strokecol}{rgb}{1.0,1.0,1.0};
        \pgfsetstrokecolor{strokecol}
        \definecolor{fillcol}{rgb}{1.0,1.0,1.0};
        \pgfsetfillcolor{fillcol}
        \filldraw (0.0bp,0.0bp) -- (0.0bp,468.0bp) -- (401.38bp,468.0bp) -- (401.38bp,0.0bp) -- cycle;
      \end{scope}
      \begin{scope}
        \pgfsetstrokecolor{black}
        \definecolor{strokecol}{rgb}{1.0,1.0,1.0};
        \pgfsetstrokecolor{strokecol}
        \definecolor{fillcol}{rgb}{1.0,1.0,1.0};
        \pgfsetfillcolor{fillcol}
        \filldraw (0.0bp,0.0bp) -- (0.0bp,468.0bp) -- (401.38bp,468.0bp) -- (401.38bp,0.0bp) -- cycle;
      \end{scope}
      \begin{scope}
        \pgfsetstrokecolor{black}
        \definecolor{strokecol}{rgb}{1.0,1.0,1.0};
        \pgfsetstrokecolor{strokecol}
        \definecolor{fillcol}{rgb}{1.0,1.0,1.0};
        \pgfsetfillcolor{fillcol}
        \filldraw (0.0bp,0.0bp) -- (0.0bp,468.0bp) -- (401.38bp,468.0bp) -- (401.38bp,0.0bp) -- cycle;
      \end{scope}
      \begin{scope}
        \pgfsetstrokecolor{black}
        \definecolor{strokecol}{rgb}{1.0,1.0,1.0};
        \pgfsetstrokecolor{strokecol}
        \definecolor{fillcol}{rgb}{1.0,1.0,1.0};
        \pgfsetfillcolor{fillcol}
        \filldraw (0.0bp,0.0bp) -- (0.0bp,468.0bp) -- (401.38bp,468.0bp) -- (401.38bp,0.0bp) -- cycle;
      \end{scope}
      \begin{scope}
        \pgfsetstrokecolor{black}
        \definecolor{strokecol}{rgb}{1.0,1.0,1.0};
        \pgfsetstrokecolor{strokecol}
        \definecolor{fillcol}{rgb}{1.0,1.0,1.0};
        \pgfsetfillcolor{fillcol}
        \filldraw (0.0bp,0.0bp) -- (0.0bp,468.0bp) -- (401.38bp,468.0bp) -- (401.38bp,0.0bp) -- cycle;
      \end{scope}
      \begin{scope}
        \pgfsetstrokecolor{black}
        \definecolor{strokecol}{rgb}{1.0,1.0,1.0};
        \pgfsetstrokecolor{strokecol}
        \definecolor{fillcol}{rgb}{1.0,1.0,1.0};
        \pgfsetfillcolor{fillcol}
        \filldraw (0.0bp,0.0bp) -- (0.0bp,468.0bp) -- (401.38bp,468.0bp) -- (401.38bp,0.0bp) -- cycle;
      \end{scope}
      \begin{scope}
        \pgfsetstrokecolor{black}
        \definecolor{strokecol}{rgb}{1.0,1.0,1.0};
        \pgfsetstrokecolor{strokecol}
        \definecolor{fillcol}{rgb}{1.0,1.0,1.0};
        \pgfsetfillcolor{fillcol}
        \filldraw (0.0bp,0.0bp) -- (0.0bp,468.0bp) -- (401.38bp,468.0bp) -- (401.38bp,0.0bp) -- cycle;
      \end{scope}
        \pgfsetcolor{black}
        \draw [] (357.0bp,215.83bp) .. controls (357.0bp,205.0bp) and (357.0bp,191.29bp)  .. (357.0bp,180.41bp);
        \draw [] (214.01bp,360.94bp) .. controls (219.98bp,349.63bp) and (227.79bp,334.83bp)  .. (233.8bp,323.44bp);
        \draw [dotted] (169.4bp,215.89bp) .. controls (172.25bp,185.29bp) and (180.19bp,121.82bp)  .. (200.0bp,72.0bp) .. controls (205.08bp,59.235bp) and (213.21bp,46.166bp)  .. (220.25bp,36.084bp);
        \draw [solid] (205.0bp,431.83bp) .. controls (205.0bp,421.0bp) and (205.0bp,407.29bp)  .. (205.0bp,396.41bp);
        \draw [dashed] (243.0bp,287.83bp) .. controls (243.0bp,277.0bp) and (243.0bp,263.29bp)  .. (243.0bp,252.41bp);
        \draw [] (243.0bp,215.83bp) .. controls (243.0bp,205.0bp) and (243.0bp,191.29bp)  .. (243.0bp,180.41bp);
        \draw [dotted] (153.1bp,290.42bp) .. controls (144.53bp,280.33bp) and (134.57bp,266.43bp)  .. (130.0bp,252.0bp) .. controls (125.17bp,236.75bp) and (127.1bp,231.73bp)  .. (130.0bp,216.0bp) .. controls (142.43bp,148.63bp) and (149.14bp,130.37bp)  .. (185.0bp,72.0bp) .. controls (193.08bp,58.84bp) and (204.59bp,45.959bp)  .. (214.42bp,36.08bp);
        \draw [] (237.28bp,144.34bp) .. controls (232.74bp,116.86bp) and (229.42bp,64.105bp)  .. (230.47bp,36.256bp);
        \draw [dotted] (246.52bp,144.05bp) .. controls (247.6bp,116.46bp) and (244.32bp,63.903bp)  .. (239.8bp,36.173bp);
        \draw [dashed] (344.07bp,145.66bp) .. controls (334.74bp,133.88bp) and (322.2bp,118.04bp)  .. (312.88bp,106.27bp);
        \draw [] (281.46bp,75.775bp) .. controls (269.51bp,64.382bp) and (254.65bp,48.328bp)  .. (244.59bp,36.034bp);
        \draw [dotted] (290.2bp,72.937bp) .. controls (281.23bp,61.875bp) and (268.07bp,47.479bp)  .. (256.64bp,36.199bp);
        \draw [dotted] (371.9bp,218.42bp) .. controls (380.47bp,208.33bp) and (390.43bp,194.43bp)  .. (395.0bp,180.0bp) .. controls (410.23bp,131.9bp) and (396.52bp,106.82bp)  .. (360.0bp,72.0bp) .. controls (332.1bp,45.394bp) and (289.05bp,30.837bp)  .. (261.2bp,23.734bp);
        \draw [dashed] (196.04bp,360.57bp) .. controls (190.23bp,349.25bp) and (182.68bp,334.56bp)  .. (176.87bp,323.27bp);
        \draw [] (357.55bp,143.92bp) .. controls (357.23bp,124.3bp) and (353.95bp,92.977bp)  .. (338.0bp,72.0bp) .. controls (319.04bp,47.058bp) and (285.26bp,32.583bp)  .. (261.34bp,25.008bp);
        \draw [] (191.04bp,223.07bp) .. controls (223.98bp,207.43bp) and (280.62bp,180.48bp)  .. (281.0bp,180.0bp) .. controls (297.29bp,159.29bp) and (300.47bp,127.89bp)  .. (300.68bp,108.18bp);
        \draw [dotted] (228.1bp,218.42bp) .. controls (219.53bp,208.33bp) and (209.57bp,194.43bp)  .. (205.0bp,180.0bp) .. controls (188.96bp,129.33bp) and (211.54bp,66.774bp)  .. (225.08bp,36.349bp);
        \draw [] (263.38bp,293.13bp) .. controls (284.13bp,280.02bp) and (316.11bp,259.83bp)  .. (336.8bp,246.76bp);
        \draw [] (168.0bp,287.83bp) .. controls (168.0bp,277.0bp) and (168.0bp,263.29bp)  .. (168.0bp,252.41bp);
      \begin{scope}
        \definecolor{strokecol}{rgb}{0.0,0.0,0.0};
        \pgfsetstrokecolor{strokecol}
        \draw (232.0bp,468.0bp) -- (178.0bp,468.0bp) -- (178.0bp,432.0bp) -- (232.0bp,432.0bp) -- cycle;
        \draw (205.0bp,450.0bp) node {$p_{ae}$};
      \end{scope}
      \begin{scope}
        \definecolor{strokecol}{rgb}{0.0,0.0,0.0};
        \pgfsetstrokecolor{strokecol}
        \draw (261.0bp,36.0bp) -- (207.0bp,36.0bp) -- (207.0bp,0.0bp) -- (261.0bp,0.0bp) -- cycle;
        \draw (234.0bp,18.0bp) node {1};
      \end{scope}
      \begin{scope}
        \definecolor{strokecol}{rgb}{0.0,0.0,0.0};
        \pgfsetstrokecolor{strokecol}
        \draw (205.0bp,378.0bp) ellipse (27.9bp and 18.0bp);
        \draw (205.0bp,378.0bp) node {$e_{bc}$};
      \end{scope}
      \begin{scope}
        \definecolor{strokecol}{rgb}{0.0,0.0,0.0};
        \pgfsetstrokecolor{strokecol}
        \draw (243.0bp,234.0bp) ellipse (27.9bp and 18.0bp);
        \draw (243.0bp,234.0bp) node {$e_{ab}$};
      \end{scope}
      \begin{scope}
        \definecolor{strokecol}{rgb}{0.0,0.0,0.0};
        \pgfsetstrokecolor{strokecol}
        \draw (243.0bp,306.0bp) ellipse (28.7bp and 18.0bp);
        \draw (243.0bp,306.0bp) node {$e_{bd}$};
      \end{scope}
      \begin{scope}
        \definecolor{strokecol}{rgb}{0.0,0.0,0.0};
        \pgfsetstrokecolor{strokecol}
        \draw (357.0bp,162.0bp) ellipse (28.7bp and 18.0bp);
        \draw (357.0bp,162.0bp) node {$e_{ce}$};
      \end{scope}
      \begin{scope}
        \definecolor{strokecol}{rgb}{0.0,0.0,0.0};
        \pgfsetstrokecolor{strokecol}
      \end{scope}
      \begin{scope}
        \definecolor{strokecol}{rgb}{0.0,0.0,0.0};
        \pgfsetstrokecolor{strokecol}
        \draw (300.0bp,90.0bp) ellipse (28.7bp and 18.0bp);
        \draw (300.0bp,90.0bp) node {$e_{de}$};
      \end{scope}
      \begin{scope}
        \definecolor{strokecol}{rgb}{0.0,0.0,0.0};
        \pgfsetstrokecolor{strokecol}
        \draw (168.0bp,306.0bp) ellipse (28.7bp and 18.0bp);
        \draw (168.0bp,306.0bp) node {$e_{bd}$};
      \end{scope}
      \begin{scope}
        \definecolor{strokecol}{rgb}{0.0,0.0,0.0};
        \pgfsetstrokecolor{strokecol}
      \end{scope}
      \begin{scope}
        \definecolor{strokecol}{rgb}{0.0,0.0,0.0};
        \pgfsetstrokecolor{strokecol}
        \draw (243.0bp,162.0bp) ellipse (28.7bp and 18.0bp);
        \draw (243.0bp,162.0bp) node {$e_{ce}$};
      \end{scope}
      \begin{scope}
        \definecolor{strokecol}{rgb}{0.0,0.0,0.0};
        \pgfsetstrokecolor{strokecol}
      \end{scope}
      \begin{scope}
        \definecolor{strokecol}{rgb}{0.0,0.0,0.0};
        \pgfsetstrokecolor{strokecol}
        \draw (168.0bp,234.0bp) ellipse (28.7bp and 18.0bp);
        \draw (168.0bp,234.0bp) node {$e_{ab}$};
      \end{scope}
      \begin{scope}
        \definecolor{strokecol}{rgb}{0.0,0.0,0.0};
        \pgfsetstrokecolor{strokecol}
      \end{scope}
      \begin{scope}
        \definecolor{strokecol}{rgb}{0.0,0.0,0.0};
        \pgfsetstrokecolor{strokecol}
        \draw (357.0bp,234.0bp) ellipse (28.7bp and 18.0bp);
        \draw (357.0bp,234.0bp) node {$e_{ab}$};
      \end{scope}
      \begin{scope}
        \definecolor{strokecol}{rgb}{0.0,0.0,0.0};
        \pgfsetstrokecolor{strokecol}
      \end{scope}
      \end{tikzpicture}
\caption{BDD for the program shown in Fig.~\ref{fig:program}, where a dashed line represents a 0-edge, a solid line the 1-edge, and a dotted line the 0-complemented edge.}
\label{fig:bdd_prog}
\end{figure}



The main idea is that a POLP induces a (non) linear function $f: \mathbb{R}^n \to \mathbb{R}$, where $n$ is equal to the number of optimizable facts. 
In the program shown in Fig.~\ref{fig:program}, $n = 2$.
Only if $n = 1$ the function is linear since there is only one variable involved (with $n > 1$ we have a joint probability distribution, so an equation with several variables multiplied together). 
Otherwise, there are several variables multiplied together.
Consider Fig.~\ref{fig:bdd_prog} that represents the BDD for the query \texttt{path(a,e)}, denoted with $p_{ae}$, where $e_{xy}$ stands for \verb|edge(x,y)|.
There are three paths with probability greater than 0.
The induced function is the sum of the functions induced by the three paths.
For example, the leftmost path that goes to terminal 1 with a regular arc induces the following equation: $f(e_{bc},e_{bd}) = (1-e_{bc}) \cdot e_{bd} \cdot 0.72$, where $0.72 = e_{ab} \cdot e_{de}$.

To solve the optimization problem, the following operations are performed: first, the BDD for the query is built, then paths and their probabilities are computed using Algorithm~\ref{alg:paths}. 
It goes as follows: first, the BDD is reordered to have the optimizable variables first.
Then, the function \textsc{PathsProbRec} is recursively called until a node associated with a not-optimizable variable is found (including terminal nodes).
From there, to compute probabilities, function \textsc{Prob} presented in~\cite{raedt07problog} is called and an empty path list and the computed probability are returned.
Essentially, function \textsc{Prob} recursively analyzes the BDD until the terminal is found, and, from there, probabilities are computed with a dynamic programming algorithm. 
It is reported in Algorithm~\ref{alg:bdd_prob} for clarity. 
When the node is associated with an optimizable fact, the paths computed at the true and false child are extended with the current node index, provided that the path has probability greater than 0.
To speed up computation, already computed paths as well as already computed probabilities (used in function \textsc{Prob}), are stored in a table: in this way, when an already encountered node is found, the paths and the probabilities are retrieved from the memory, rather than recomputed.

From the list of paths, where optimizable variables are kept symbolically, we obtain an equation, that we call it \textit{query equation} in this way: we multiply together all the nodes $n_i$ of the paths and their probability for each path in the list of paths (if the node is selected as 0, we multiply by $1 - n_i$) and add all the results.
So, for example, for Fig.~\ref{fig:bdd_prog}, the resulting list of paths obtained by Algorithm~\ref{alg:paths} is $[[0.774,[[e_{bd},1],[e_{bc},1]]] , [ 0.27, [ [e_{bd},0], [e_{bc},1] ] ], [0.72, [ [e_{bd},1], [e_{bc},0] ]] ]$, where $0.774 = e_{ab} \cdot (e_{ce} + (1 - e_{ce}) \cdot e_{de})$, $0.72 = e_{ab} \cdot e_{de}$, and $0.27 = e_{ab} \cdot e_{ce}$ (obtained by following the correspondent paths on the BDD), which represents the equation $e_{bd} \cdot e_{bc} \cdot 0.774 + (1 - e_{bd}) \cdot e_{bc} \cdot 0.27 + e_{bd} \cdot (1 - e_{bc}) \cdot 0.72$.
The symbolic equation is then simplified and substituted in the constraint(s) involving the query.
This equation, along with all constraints specified by the user, are added to the (non-linear) optimization problem and passed to a non-linear optimization solver. 
Finally, once the optimal value(s) are calculated, the probability of the query is computed by evaluating the query equation, where the symbolical optimizable variables are substituted by their optimal values. 
The whole process is reported in Algorithm~\ref{alg:solve}.
If we consider the program shown in Figure~\ref{fig:program}, one of the possible solutions is given by $f(0.6352,0.7352) = 1.3704$, which yields a probability of the query of 0.6.

The complexity of answering probabilistic queries is, in general, \#P-complete~\cite{koller2009probabilistic}. 
The construction of the BDD starting from the derivations of the program is as well \#P-complete, and so the solution of the Probabilistic Optimizable Problem is, at least, in that class.
Finding the optimal order of the variables that minimizes the size of a BDD (and thus reduces the size of the symbolic equation) is a NP-complete task~\cite{DBLP:journals/tc/BolligW96}, so it is generally infeasible.
Moreover, if we choose this approach, every time we reorder the BDD we need to traverse it again to extract a new equation, and check whether it is more compact than the previously obtained. 
Even if the traversal of a BDD can be performed polynomially (we visit every node only once), the number of possible combinations of variables is too large to check exhaustively.
For these reasons, we decided to extract the equation directly from the BDD after having reordered it by moving nodes that correspond to optimizable facts first.
The reorder can be performed polynomially in the size of the BDD by swapping adjacent variables~\cite{jiang2017variable}.

The reordering of the BDD is crucial since it allows to directly use function \textsc{Prob} and obtain a more compact symbolic representation, where the combinations of the optimizable variables are multiplied by a single numerical value.
Similarly, simplification of the query equation is fundamental, since it greatly reduces its size, resulting in faster computation. 
Consider, for example, the equation represented by the BDD shown in Fig.~\ref{fig:bdd_prog} written before and reported here for clarity: $e_{bd} \cdot e_{bc} \cdot 0.774 + (1 - e_{bd}) \cdot e_{bc} \cdot 0.27 + e_{bd} \cdot (1 - e_{bc}) \cdot 0.72$.
After the simplification, it is reduced as: $-0.216 \cdot e_{bc} \cdot e_{bd}+ 0.27 \cdot e_{bc} + 0.72 \cdot e_{bd}$ (the number of operations is now 6, previously was 10).
Moreover, this process may also reduce the impact of the BDD structure, since the order of the variables in the BDD also determines its size: several BDDs can represent the same function, but some of these BDDs may not be optimal in terms of compactness. 
If we directly utilize the query equation extracted by Algorithm~\ref{alg:paths}, without simplifying the result, we may obtain a long equation that involves more computations (multiplications and summations) than the strict necessary.
Since the query equation can be called several times during the solution of the optimization problem, not simplifying it can largely increase the overall execution time.

\begin{algorithm}[t]
\begin{scriptsize}
\caption{Function \textsc{OptimizeProb}: optimization of probability of random variables.} 
\label{alg:solve}
\begin{algorithmic}[1]

\Function{OptimizeProb}{query,objective,constraintsList,algorithm}
\State root $\gets$ Compute the BDD for the query
\State paths $\gets$ \Call{PathsProb}{root}
\State query equation $\gets$ convert paths into symbolic equation
\State Simplify query equation
\State Replace constraints from constraintsList involving the query with query equation
\State assignments $\gets$ Call the non-linear optimization solver with [objective, constraintsList, algorithm]
\State prob $\gets$ evaluate(query equation $\mid$ assignments)
\State \Return [assignments, prob]
\EndFunction
\end{algorithmic}
\end{scriptsize}
\end{algorithm}


\begin{algorithm}[t]
\begin{scriptsize}
\caption{Function \textsc{PathsProb}: computation of all the paths of a BDD and of their probability.} 
\label{alg:abd}
\begin{algorithmic}[1]
\Function{PathsProb}{$root$}
\State $root' \gets \Call{Reorder}{root}$ \label{alg_abd_reorder} \Comment{BDD reordering}
\State $\TablePaths \gets \emptyset$
\State $\TableProb \gets \emptyset$
\If{$root'.comp$}
  \State $comp \gets true$
\Else
  \State $comp \gets false$
\EndIf
\State \Return \Call{PathsProbRec}{$root',comp$}
\EndFunction
\Function{PathsProbRec}{$node,comp$}
\State $comp \gets node.comp \oplus comp$
\If{$var(node)$ is not associated to an optimizable fact}
  \State $p \gets $\Call{Prob}{$node$} \label{alg_abd_call_prob} \Comment{Call to \Call{prob}{}}
  \If{$comp$}
    \State $\Res \gets [1-p,[]]$
  \Else
    \State $\Res \gets [p,[]]$
  \EndIf
\Else
  \If{$\TablePaths(node.index) \neq \emptyset$}
    \State\Return $\TablePaths(node.index)$
  \Else
    \State $Lp_0 \gets$ \Call{PathsProbRec}{$child_0(node),\comp$}
    \State $Lp_1 \gets$ \Call{PathsProbRec}{$child_1(node),\comp$}
    \State $\Res \gets []$
    \ForAll{$path \in Lp_0$}
      \If{$path.prob > 0$}
        \State $\Res \gets Res \cup \{ path \cup [node.index,0]\}$
      \EndIf
    \EndFor
    \ForAll{$path \in Lp_1$}
      \If{$path.prob > 0$}
        \State $\Res \gets Res \cup \{ path \cup [node.index,1]\}$
      \EndIf
    \EndFor
  \EndIf
  \State Add $node.index \rightarrow \Res$ to $\TablePaths$
\EndIf
  \State \Return $\Res$
\EndFunction
\end{algorithmic}
\label{alg:paths}
\end{scriptsize}
\end{algorithm}

\begin{algorithm}
  \begin{scriptsize}
  \caption{Function Prob: computation of the probability of a BDD.} 
  \label{alg:bdd_prob}
  \begin{algorithmic}[1]
  \Function{Prob}{$node$}
  \If{$node$ is a terminal}
    \State \Return $1$
  \Else
    \If{$\TableProb(node.pointer)\neq null$}
      \State\Return $\TableProb(node)$
    \Else
      \State $p0\gets$\Call{Prob}{$child_0(node)$}
      \State $p1\gets$\Call{Prob}{$child_1(node)$}
      \If{$child_0(node).comp$}
        \State $p0\gets (1-p0)$
      \EndIf
      \State Let $\pi$ be the probability of being true of $var(node)$
      \State $Res\gets p1\cdot \pi+p0\cdot (1-\pi)$
      \State Add $node.pointer\rightarrow Res$ to $\TableProb$
      \State\Return $Res$
    \EndIf
  \EndIf
  \EndFunction
  \end{algorithmic}
  \end{scriptsize}
\end{algorithm}  
  
\section{Experiments}
\label{sec:exp}
We implemented the algorithm presented before\footnote{Source code available at: \url{https://bitbucket.org/machinelearningunife/polp_experiments}} using C and exploiting the library for nonlinear optimization NLopt~\cite{nlopt} to solve the optimization problem and the library CUDD~\cite{somenzi2015} for the operations on BDD.
The Probabilistic Logic Part is managed by SWI-Prolog~\cite{DBLP:journals/tplp/WielemakerSTL12} version 8.3.15.
The simplification of the query equation is performed with the function \verb|simplify| from Python SymPy package~\cite{meurer2017sympy}: guided by some heuristics, it iteratively applies some simplifications to the equation to reduce its size. However, in general, there is no guarantee that the minimal size is found.

To test our algorithm on real world scenarios described in Section~\ref{sec:intro}, we selected from~\cite{rossi2015network} several dataset characterized by different structures, among them: social network, collaboration network, road connections, and power network.
We pre-process the data by converting the matrix in a set of facts of the form \texttt{edge(a,b)} (eventually prepending the functor \texttt{optimizable} and adding a probability range or value) to represent a connection between nodes \texttt{a} and \texttt{b}.
We choose these datasets since they represent several real world networks, linked to the motivations listed in Section~\ref{sec:intro}. 
We selected to test the implementation on, among other, social networks, collaboration networks, road networks, and power networks.
For all of these, the goal is to constrain the probability of the paths between a random source (\texttt{Source}) and a random destination (\texttt{Dest}), provided that the path between these two nodes exists, to be greater than $0.8$, while minimizing the sum of the probabilities of optimizable edges.
Note that we minimized the sum of the probabilities of all the optimizable facts, not only the ones involved in the query, since it may be difficult to spot which facts are involved in it.
For example, if you want to compute all the paths from a source to a destination, it is difficult to say which edges will be involved, without running the program.
The number of optimizable facts (edges) is randomly set to 50\% of the total number of edges, and the values lower and upper bounds of the probability of optimizable facts are respectively 0.001 and 0.999.
The remaining facts are probabilistic facts with probability $0.5$.
Results are averages of 10 runs with randomly chosen source and destination.
Facts (both optimizable and probabilistic) are of the form \texttt{edge/2}, and the query is \texttt{path(Source,Dest) > 0.8}, where \texttt{path/2} is the predicate described above (Fig~\ref{fig:program}).

In a second set of experiments, we generate complete graphs of increasing size.
As before, the goal is to minimize the sum of the optimizable facts while the probability of the path between 1 and $N$ (where $N$ is the size of the graph) must be above 0.8.
The subdivision between optimizable and probabilistic facts is the same as before (50-50) as well as the structure of the facts and the query (\texttt{edge/2} facts, \texttt{path(1,N) > 0.8}).
For this experiment, the graph size (and thus the number of optimizable facts) is notably smaller than the previous dataset.
However, the solution of the optimization problem is harder, since the graph is fully connected, so there are many paths from the root of the BDD to the terminal, resulting in a very long and complex constraint function.

For all the experiments, we tested three local gradient-based optimization algorithms available in NLopt~\cite{nlopt}: two based on conservative convexseparable approximations~\cite{svanberg2002ccsa}, denoted with MMA and CCSAQ, and one based on sequential quadratic programming~\cite{Dieter1994slsqp}, denoted with SLSQP. 

Experiments were conducted on a cluster\footnote{\url{https://www.fe.infn.it/coka/doku.php}} with Intel\textsuperscript{\textregistered} Xeon\textsuperscript{\textregistered} E5-2630v3 running at 2.40 GHz.
We set the tolerance to $10^{-5}$, that is, the optimization process stops when the variation of the objective function is less than this value.
The available memory for the execution has been set to 8GB.
Execution times are computed with the SWI-Prolog predicate \verb|statistics/2| with the keyword \verb|runtime|.
The maximum execution time for each experiment was 8 hours.

Results are shown in Table~\ref{tab:results} and Table~\ref{tab:complete_results}, which report the average overall execution time (BDD generation, simplification and optimization) and the average probability for all three algorithms.
For the first set of experiments, we also tabled the standard deviations of the 10 averaged probabilities and the number of vertices and edges for each dataset.
We marked in bold the best algorithm for every experiment.
In Table~\ref{tab:results}, the dataset bio stands for bio-DM-LC, ca for ca-netscience, E60 for ENZYMES\_g60, IIP for internet-industry-partnerships, p949 for power-494-bus, p669 for power-662-bus, rtf for reptilia-tortoise-fi-2008, rc for road-chesapeake, rt for rt-retweet, soc for soc-tribes, and web for webkb-wisc (all available at~\cite{rossi2015network}).

\begin{table*}[htb]
  \centering
  \begin{tabular}{ c | c | c || c | c | c || c | c | c || c | c | c|  }
    \cline{2-12} & \multicolumn{2}{c||}{Features} & \multicolumn{3}{c||}{Time (s)} & \multicolumn{3}{c||}{Probability}  & \multicolumn{3}{c|}{StdDev (Prob)} \\ \cline{1-12}

    Dataset & \texttt{|V|} & \texttt{|E|} & C & M & S & C & M & S & C & M & S \\
    \hline			
    bio     & 658 & 1129  & 2595 & 4934 & \textbf{1072}  & 0.850 & 0.823 & \textbf{0.800} & 0.065 &	0.059 & \textbf{0.004} \\
    ca & 379 & 914   & 2859 & 2137 &	\textbf{387}   & 0.821 & 0.823 & \textbf{0.816} & \textbf{0.023} & 0.031 & 0.030 \\
    DD244         & 291 & 882   & 2070 & 2355 & \textbf{521} & 0.822 &	0.815 & \textbf{0.800}	& 0.037	& 0.030 & \textbf{0.000} \\
    E60 & 10   & 36    & \textbf{25}  & 30    & 37    & \textbf{0.887} & \textbf{0.887} & \textbf{0.887} & 0.068 & \textbf{0.065} & \textbf{0.065} \\
    IIP & 219 & 613 & 1079 & 690 & \textbf{209} & 0.802 & \textbf{0.755} & 0.801 & \textbf{0.004} & 0.190 & \textbf{0.004} \\
    p494 & 494 & 586 & 3026 & 3898 & \textbf{1052}   & 0.832 &	0.820 &	\textbf{0.800} & 0.041 &	0.029 &	\textbf{0.000} \\
    p662 & 662 & 906 & 16167 & 7990 & \textbf{2292}  & 0.802 & 0.824 & \textbf{0.800} & 0.003 &	0.020 &	\textbf{0.000} \\
    rtf & 283 & 418 & 271 & 211 & \textbf{48} & 0.817 & 0.826 & \textbf{0.800} & 0.029 & 0.026 & \textbf{0.000} \\
    rc & 39 & 170 & 47 & 35 & \textbf{7} & 0.837 & 0.864 & \textbf{0.829} & 0.063 & 0.065 & \textbf{0.060} \\
    rt & 97 & 117 & 21 & 13 & \textbf{3} & 0.827 & 0.840 & \textbf{0.802} & 0.058 & 0.056 & \textbf{0.004} \\
    soc & 16 & 58 & 133 & 133 & \textbf{11} & \textbf{0.842} & \textbf{0.842} & \textbf{0.842} & \textbf{0.057} & \textbf{0.057} & \textbf{0.057} \\
    web & 265 & 530 & 829 & 712 & \textbf{146} & 0.821 & 0.821 & \textbf{0.801} & 0.057 & 0.044 & \textbf{0.003} \\
    \hline  
  \end{tabular}
  \caption{Results for networks experiments. C, M and S stand respectively for CCSAQ, MMA, and SLSQP algorithms. \texttt{|V|} and \texttt{|E|} are number of vertices and edges.}
  \label{tab:results}
\end{table*}

\begin{table}[htb]
  \centering
  \begin{tabular}{  c  c c  c  c  c  c  }
  N & C (s) & M (s) & S (s) & C (prob) & M (prob) & S (prob) \\
  \hline
  3  & 1.6 & 1.7 & \textbf{0.4} & 0.800 & 0.800 & 0.800 \\
  4  & 8.2 & 2.7 & \textbf{0.4} & 0.818 & 0.800 & 0.800 \\
  5  & 4.5 & 4.7 & \textbf{0.9} & 0.800 & 0.800 & 0.800 \\
  6  & 10.7 & 12.4 & \textbf{1.3} & 0.800 & 0.800 & 0.800 \\
  7  & 202.4 & 193.8 & \textbf{30.1} & 0.800 & 0.800 & 0.800 \\
  8  & 1,360.3 & 1,600.6 & \textbf{177.6} & 0.800 & 0.800 & 0.800 \\ 
  \hline
\end{tabular}
\caption{Results for complete graphs experiments.}
\label{tab:complete_results}
\end{table}

The time for the BDD construction and for the query equation extraction is an order of magnitude less than the time used for the optimization, often negligible. 
This is also the case for the simplification of the equation (even if it is a little more expensive than the BDD extraction, and sometimes, for larger graphs, it requires several seconds), so we decided to not separate the three values. 
Rather, we provide only the sum of the execution times of the three operations.

Usually, the best algorithm (fastest and more accurate in terms of difference with the lower probability, 0.8 for these experiments) is SLSQP, with an important difference in terms of both execution time and probability in most of the cases. 
Especially, for the complete graph dataset, SLSQP requires almost ten times less the time required by CCSAQ and MMA for bigger graphs.

The execution time exceeds eight hours for complete graph of size 9. 
The main reason is that nodes have high degree and so the number of paths increases exponentially, resulting in an explosion of the length of the query equation.
To see this phenomenon, consider the dataset bio-DM-LC and power-494-bus: the former has approximately 200 vertices and edges more than the latter. 
However, execution times are comparable, especially for SLSQP.
This also occurs for different networks we tested, still taken from~\cite{rossi2015network}, where some of the queries does not terminate in the time limit, such as soc-firm-hi-tech, ia-crime-moreno, lp\_adlittle, and soc-wiki-vote.

The bottleneck of our approach is, as expected, in the solution of the optimization problem, rather than the construction of the BDD representing the program or the simplification of the query equation.
One possible trivial solution to this problem is reducing the tolerance of the optimization algorithm, but at the cost of less accurate results.
Alternatively, we can try to provide more compact representations of the programs (some alternatives to BDDs exist, as discussed at the beginning of the paper), or leverage techniques from lifted inference. 
These are all directions of possible interesting future works.

  
\section{Conclusions}
\label{sec:concl}
In this paper, we introduced the definition of Probabilistic Optimizable Logic Program and proposed a simple yet effective algorithm to find the optimal probabilities of optimizable facts subject to constraints.
Empirical results show that the proposed solution is scalable also for datasets with several hundreds of vertices.
As future work, we planned to test other optimization packages and apply this approach to programs with continuous random variables~\cite{AzzRigLamMas19-AIXIA-IC,AzzRigLam21-AIJ-IJ}.

\bibliographystyle{acmtrans}
\bibliography{bibl}

\begin{thebibliography}{}

\bibitem[\protect\citeauthoryear{Antuori and Richoux}{Antuori and
  Richoux}{2019}]{antuori2019iee}
{\sc Antuori, V.} {\sc and} {\sc Richoux, F.} 2019.
\newblock Constrained optimization under uncertainty for decision-making
  problems: Application to real-time strategy games.
\newblock In {\em 2019 IEEE Congress on Evolutionary Computation (CEC)}.
  458--465.

\bibitem[\protect\citeauthoryear{Azzolini, Riguzzi, and Lamma}{Azzolini
  et~al\mbox{.}}{2021}]{AzzRigLam21-AIJ-IJ}
{\sc Azzolini, D.}, {\sc Riguzzi, F.}, {\sc and} {\sc Lamma, E.} 2021.
\newblock A semantics for hybrid probabilistic logic programs with function
  symbols.
\newblock {\em Artificial Intelligence\/}~{\em 294}, 103452.

\bibitem[\protect\citeauthoryear{Azzolini, Riguzzi, Lamma, and
  Masotti}{Azzolini et~al\mbox{.}}{2019}]{AzzRigLamMas19-AIXIA-IC}
{\sc Azzolini, D.}, {\sc Riguzzi, F.}, {\sc Lamma, E.}, {\sc and} {\sc Masotti,
  F.} 2019.
\newblock A comparison of {MCMC} sampling for probabilistic logic programming.
\newblock In {\em Proceedings of the 18th Conference of the Italian Association
  for Artificial Intelligence ({AI*IA2019}), Rende, Italy 19-22 November 2019},
  {M.~Alviano}, {G.~Greco}, {and} {F.~Scarcello}, Eds. Lecture Notes in
  Computer Science. Springer, Heidelberg, Germany.

\bibitem[\protect\citeauthoryear{Babaki, Guns, and de~Raedt}{Babaki
  et~al\mbox{.}}{2017}]{babaki2017ijcai}
{\sc Babaki, B.}, {\sc Guns, T.}, {\sc and} {\sc de~Raedt, L.} 2017.
\newblock Stochastic constraint programming with and-or branch-and-bound.
\newblock In {\em Proceedings of the Twenty-Sixth International Joint
  Conference on Artificial Intelligence, {IJCAI-17}}. 539--545.

\bibitem[\protect\citeauthoryear{Bellodi and Riguzzi}{Bellodi and
  Riguzzi}{2012}]{BelRig12-ILP11-IC}
{\sc Bellodi, E.} {\sc and} {\sc Riguzzi, F.} 2012.
\newblock Learning the structure of probabilistic logic programs.
\newblock In {\em 22nd International Conference on Inductive Logic
  Programming}, {S.~Muggleton}, {A.~Tamaddoni-Nezhad}, {and} {F.~Lisi}, Eds.
  Lecture Notes in Computer Science, vol. 7207. Springer Berlin Heidelberg,
  61--75.

\bibitem[\protect\citeauthoryear{Bollig and Wegener}{Bollig and
  Wegener}{1996}]{DBLP:journals/tc/BolligW96}
{\sc Bollig, B.} {\sc and} {\sc Wegener, I.} 1996.
\newblock Improving the variable ordering of {OBDDs} is {NP}-complete.
\newblock {\em IEEE Trans. Computers\/}~{\em 45,\/}~9, 993--1002.

\bibitem[\protect\citeauthoryear{Darwiche and Marquis}{Darwiche and
  Marquis}{2002}]{DBLP:journals/jair/DarwicheM02}
{\sc Darwiche, A.} {\sc and} {\sc Marquis, P.} 2002.
\newblock A knowledge compilation map.
\newblock {\em Journal of Artificial Intelligence Research\/}~{\em 17},
  229--264.

\bibitem[\protect\citeauthoryear{{De Raedt}, Frasconi, Kersting, and
  Muggleton}{{De Raedt} et~al\mbox{.}}{2008}]{DBLP:conf/ilp/2008p}
{\sc {De Raedt}, L.}, {\sc Frasconi, P.}, {\sc Kersting, K.}, {\sc and} {\sc
  Muggleton, S.}, Eds. 2008.
\newblock {\em Probabilistic Inductive Logic Programming}. LNCS, vol. 4911.
  Springer.

\bibitem[\protect\citeauthoryear{{De Raedt}, Kimmig, and Toivonen}{{De Raedt}
  et~al\mbox{.}}{2007}]{raedt07problog}
{\sc {De Raedt}, L.}, {\sc Kimmig, A.}, {\sc and} {\sc Toivonen, H.} 2007.
\newblock Problog: A probabilistic prolog and its application in link
  discovery.
\newblock In {\em IJCAI}, {M.~M. Veloso}, Ed. 2462--2467.

\bibitem[\protect\citeauthoryear{Fierens, {Van den Broeck}, Renkens,
  Shterionov, Gutmann, Thon, Janssens, and {De Raedt}}{Fierens
  et~al\mbox{.}}{2015}]{DBLP:journals/corr/abs-1304-6810}
{\sc Fierens, D.}, {\sc {Van den Broeck}, G.}, {\sc Renkens, J.}, {\sc
  Shterionov, D.~S.}, {\sc Gutmann, B.}, {\sc Thon, I.}, {\sc Janssens, G.},
  {\sc and} {\sc {De Raedt}, L.} 2015.
\newblock Inference and learning in probabilistic logic programs using weighted
  {Boolean} formulas.
\newblock {\em Theory and Practice of Logic Programming\/}~{\em 15,\/}~3,
  358--401.

\bibitem[\protect\citeauthoryear{Gutmann, Kimmig, Kersting, and
  De~Raedt}{Gutmann et~al\mbox{.}}{2008}]{DBLP:conf/pkdd/GutmannKKR08}
{\sc Gutmann, B.}, {\sc Kimmig, A.}, {\sc Kersting, K.}, {\sc and} {\sc
  De~Raedt, L.} 2008.
\newblock Parameter learning in probabilistic databases: A least squares
  approach.
\newblock In {\em European Conference on Machine Learning and Principles and
  Practice of Knowledge Discovery in Databases (ECMLPKDD 2008)}. Lecture Notes
  in Computer Science, vol. 5211. Springer, 473--488.

\bibitem[\protect\citeauthoryear{Gutmann, Thon, and De~Raedt}{Gutmann
  et~al\mbox{.}}{2011}]{DBLP:conf/pkdd/GutmannTR11}
{\sc Gutmann, B.}, {\sc Thon, I.}, {\sc and} {\sc De~Raedt, L.} 2011.
\newblock Learning the parameters of probabilistic logic programs from
  interpretations.
\newblock In {\em European Conference on Machine Learning and Principles and
  Practice of Knowledge Discovery in Databases (ECMLPKDD 2011)},
  {D.~Gunopulos}, {T.~Hofmann}, {D.~Malerba}, {and} {M.~Vazirgiannis}, Eds.
  Lecture Notes in Computer Science, vol. 6911. Springer, 581--596.

\bibitem[\protect\citeauthoryear{Jiang, Babar, Ciardo, Miner, and Smith}{Jiang
  et~al\mbox{.}}{2017}]{jiang2017variable}
{\sc Jiang, C.}, {\sc Babar, J.}, {\sc Ciardo, G.}, {\sc Miner, A.~S.}, {\sc
  and} {\sc Smith, B.} 2017.
\newblock Variable reordering in binary decision diagrams.
\newblock In {\em 26th International Workshop on Logic and Synthesis}. 1--8.

\bibitem[\protect\citeauthoryear{Johnson}{Johnson}{2020}]{nlopt}
{\sc Johnson, S.~G.} 2020.
\newblock The nlopt nonlinear-optimization package.

\bibitem[\protect\citeauthoryear{Kimmig, Van~den Broeck, and De~Raedt}{Kimmig
  et~al\mbox{.}}{2011}]{kimmig2011algebraic}
{\sc Kimmig, A.}, {\sc Van~den Broeck, G.}, {\sc and} {\sc De~Raedt, L.} 2011.
\newblock An algebraic prolog for reasoning about possible worlds.
\newblock In {\em Proceedings of the Twenty-Fifth AAAI Conference on Artificial
  Intelligence}. Vol.~1. AAAI Press, 209--214.

\bibitem[\protect\citeauthoryear{Koller and Friedman}{Koller and
  Friedman}{2009}]{koller2009probabilistic}
{\sc Koller, D.} {\sc and} {\sc Friedman, N.} 2009.
\newblock {\em Probabilistic Graphical Models: Principles and Techniques}.
\newblock Adaptive computation and machine learning. MIT Press, Cambridge, MA.

\bibitem[\protect\citeauthoryear{Kraft}{Kraft}{1994}]{Dieter1994slsqp}
{\sc Kraft, D.} 1994.
\newblock Algorithm 733: Tomp–fortran modules for optimal control
  calculations.
\newblock ~{\em 20,\/}~3 (Sept.), 262–281.

\bibitem[\protect\citeauthoryear{Latour, Babaki, Dries, Kimmig, Van~den Broeck,
  and Nijssen}{Latour et~al\mbox{.}}{2017}]{10.1007/978-3-319-66158-2_32}
{\sc Latour, A. L.~D.}, {\sc Babaki, B.}, {\sc Dries, A.}, {\sc Kimmig, A.},
  {\sc Van~den Broeck, G.}, {\sc and} {\sc Nijssen, S.} 2017.
\newblock Combining stochastic constraint optimization and probabilistic
  programming.
\newblock In {\em Principles and Practice of Constraint Programming}, {J.~C.
  Beck}, Ed. Springer International Publishing, Cham, 495--511.

\bibitem[\protect\citeauthoryear{Lombardi and Milano}{Lombardi and
  Milano}{2010}]{lombardi2010aij}
{\sc Lombardi, M.} {\sc and} {\sc Milano, M.} 2010.
\newblock Allocation and scheduling of conditional task graphs.
\newblock {\em Artificial Intelligence\/}~{\em 174,\/}~7, 500--529.

\bibitem[\protect\citeauthoryear{Meurer, Smith, Paprocki, \v{C}ert\'{i}k,
  Kirpichev, Rocklin, Kumar, Ivanov, Moore, Singh, Rathnayake, Vig, Granger,
  Muller, Bonazzi, Gupta, Vats, Johansson, Pedregosa, Curry, Terrel,
  Rou\v{c}ka, Saboo, Fernando, Kulal, Cimrman, and Scopatz}{Meurer
  et~al\mbox{.}}{2017}]{meurer2017sympy}
{\sc Meurer, A.}, {\sc Smith, C.~P.}, {\sc Paprocki, M.}, {\sc \v{C}ert\'{i}k,
  O.}, {\sc Kirpichev, S.~B.}, {\sc Rocklin, M.}, {\sc Kumar, A.}, {\sc Ivanov,
  S.}, {\sc Moore, J.~K.}, {\sc Singh, S.}, {\sc Rathnayake, T.}, {\sc Vig,
  S.}, {\sc Granger, B.~E.}, {\sc Muller, R.~P.}, {\sc Bonazzi, F.}, {\sc
  Gupta, H.}, {\sc Vats, S.}, {\sc Johansson, F.}, {\sc Pedregosa, F.}, {\sc
  Curry, M.~J.}, {\sc Terrel, A.~R.}, {\sc Rou\v{c}ka, v.}, {\sc Saboo, A.},
  {\sc Fernando, I.}, {\sc Kulal, S.}, {\sc Cimrman, R.}, {\sc and} {\sc
  Scopatz, A.} 2017.
\newblock Sympy: symbolic computing in python.
\newblock {\em PeerJ Computer Science\/}~{\em 3}, e103.

\bibitem[\protect\citeauthoryear{Orsini, Frasconi, and De~Raedt}{Orsini
  et~al\mbox{.}}{2017}]{orsini2017kproblog}
{\sc Orsini, F.}, {\sc Frasconi, P.}, {\sc and} {\sc De~Raedt, L.} 2017.
\newblock {kProbLog}: an algebraic prolog for machine learning.
\newblock {\em Machine Learning\/}~{\em 106,\/}~12, 1933--1969.

\bibitem[\protect\citeauthoryear{Riguzzi}{Riguzzi}{2018}]{Rig18-BK}
{\sc Riguzzi, F.} 2018.
\newblock {\em Foundations of Probabilistic Logic Programming}.
\newblock River Publishers, Gistrup, Denmark.

\bibitem[\protect\citeauthoryear{Riguzzi and Swift}{Riguzzi and
  Swift}{2010}]{RigSwi10-ICLP10-IC}
{\sc Riguzzi, F.} {\sc and} {\sc Swift, T.} 2010.
\newblock Tabling and answer subsumption for reasoning on logic programs with
  annotated disjunctions.
\newblock In {\em Technical Communications of the 26th International Conference
  on Logic Programming ({ICLP} 2010}. LIPIcs, vol.~7. Schloss Dagstuhl,
  162--171.

\bibitem[\protect\citeauthoryear{Rossi and Ahmed}{Rossi and
  Ahmed}{2015}]{rossi2015network}
{\sc Rossi, R.~A.} {\sc and} {\sc Ahmed, N.~K.} 2015.
\newblock The network data repository with interactive graph analytics and
  visualization.
\newblock In {\em AAAI}.

\bibitem[\protect\citeauthoryear{Sato}{Sato}{1995}]{DBLP:conf/iclp/Sato95}
{\sc Sato, T.} 1995.
\newblock A statistical learning method for logic programs with distribution
  semantics.
\newblock In {\em Logic Programming, Proceedings of the Twelfth International
  Conference on Logic Programming}, {L.~Sterling}, Ed. MIT Press, 715--729.

\bibitem[\protect\citeauthoryear{Sato and Kameya}{Sato and
  Kameya}{2001}]{DBLP:journals/jair/SatoK01}
{\sc Sato, T.} {\sc and} {\sc Kameya, Y.} 2001.
\newblock Parameter learning of logic programs for symbolic-statistical
  modeling.
\newblock {\em Journal of Artificial Intelligence Research\/}~{\em 15},
  391--454.

\bibitem[\protect\citeauthoryear{Somenzi}{Somenzi}{2015}]{somenzi2015}
{\sc Somenzi, F.} 2015.
\newblock {\em CUDD: CU Decision Diagram Package Release 3.0.0}.
\newblock University of Colorado.

\bibitem[\protect\citeauthoryear{Svanberg}{Svanberg}{2002}]{svanberg2002ccsa}
{\sc Svanberg, K.} 2002.
\newblock A class of globally convergent optimization methods based on
  conservative convex separable approximations.
\newblock {\em SIAM Journal on Optimization\/}, 555--573.

\bibitem[\protect\citeauthoryear{Van~den Broeck, Thon, van Otterlo, and
  De~Raedt}{Van~den Broeck et~al\mbox{.}}{2010}]{DBLP:conf/aaai/BroeckTOR10}
{\sc Van~den Broeck, G.}, {\sc Thon, I.}, {\sc van Otterlo, M.}, {\sc and} {\sc
  De~Raedt, L.} 2010.
\newblock {DTProbLog}: A decision-theoretic probabilistic {Prolog}.
\newblock In {\em Proceedings of the Twenty-Fourth AAAI Conference on
  Artificial Intelligence}, {M.~Fox} {and} {D.~Poole}, Eds. AAAI Press,
  1217--1222.

\bibitem[\protect\citeauthoryear{Vennekens, Verbaeten, and
  Bruynooghe}{Vennekens et~al\mbox{.}}{2004}]{VenVer04-ICLP04-IC}
{\sc Vennekens, J.}, {\sc Verbaeten, S.}, {\sc and} {\sc Bruynooghe, M.} 2004.
\newblock Logic programs with annotated disjunctions.
\newblock In {\em 20th International Conference on Logic Programming (ICLP
  2004)}, {B.~Demoen} {and} {V.~Lifschitz}, Eds. Lecture Notes in Computer
  Science, vol. 3131. Springer, 431--445.

\bibitem[\protect\citeauthoryear{Walsh}{Walsh}{2002}]{walsh2002}
{\sc Walsh, T.} 2002.
\newblock Stochastic constraint programming.
\newblock {\em Proceedings of the 15th European Conference on Artificial
  Intelligence\/}~{\em 1}, 111--115.

\bibitem[\protect\citeauthoryear{Wielemaker, Schrijvers, Triska, and
  Lager}{Wielemaker et~al\mbox{.}}{2012}]{DBLP:journals/tplp/WielemakerSTL12}
{\sc Wielemaker, J.}, {\sc Schrijvers, T.}, {\sc Triska, M.}, {\sc and} {\sc
  Lager, T.} 2012.
\newblock {SWI-Prolog}.
\newblock {\em Theory and Practice of Logic Programming\/}~{\em 12,\/}~1-2,
  67--96.

\end{thebibliography}

\end{document}